\documentstyle{article}
\begin{document}

\title{%
The Effect of a Small Mixing Angle in the Atmospheric Neutrinos
\footnote{Talk presented at the 1st Workshop on {\it Neutrino Oscillations 
and their Origin}, February 11-13, 2000, Fujiyoshida, Japan}
}

\author{%
      Shoichi Midorikawa\footnote{Email: midori@aomori-u.ac.jp}
\\
{\it  Faculty of Engineering,
 Aomori University, Japan. midori@aomori-u.ac.jp, 
}\\
}

\date{}

\maketitle

\section*{Abstract}

The effect of matter enhanced neutrino oscillations on atmosheric 
neutrinos is investigated systematically in the framework of one mass 
dominant model of three neutrinos.
The resonance conditions of neutrino crossing the earth are determined
by the three parameters, namely, the zenith angle, ${\Delta m^2/E}$, and 
the mixing angle ${\theta_3}$ of the electron neutrinos with tau neutrinos.  
The values of the triplet under the resonance is found numerically.   

\section{Introduction}

It was almost ten years ago that the atmospheric neutrino anomaly\cite{Kam88} 
could be explained by the oscillations of muon neutrinos into tau neutrinos
\cite{Learned88}. 
This interpretation has been confirmed by the observation of the zenith angle 
dependence of the neutrino fluxes by the SuperKamiokande\cite{Super-K98}.
While the zenith angle dependence of the electron-like events is consistent
with the theoretical predictions\cite{Barr89}\cite{Honda90}, that of the 
muon-like events disagree with the theory, especially for the up-going events.
     
The neutrino state with flavor basis and that of mass basis is related to 
each other by the mixing matrix as follows, 
\begin{equation}
\pmatrix{
\nu_e  \cr
\nu_{\mu} \cr
\nu_{\tau} \cr}
=
\pmatrix{
c_1c_3 & s_1c_3 &  s_3 \cr
- s_1c_2 - c_1s_2s_3 & c_1c_2 - s_1s_2s_3 & s_2c_3 \cr
s_1s_2 - c_1c_2s_3 & - c_1s_2 - s_1c_2s_3 & c_2c_3 \cr} \\
\pmatrix{
\nu_1  \cr
\nu_2  \cr
\nu_3  \cr}, 
\end{equation}
where
${
s_i = \sin \theta_i, \  c_i =\cos \theta_i \quad {\rm for} \ i =1,2, 
\ {\rm and}\  3.
}$
The Super-Kamiokande shows that 
\begin{equation}
P(\nu_\mu \rightarrow \nu_\mu) + P(\nu_\mu \rightarrow \nu_\tau) \simeq 1.
\label{eq:mu-tau_osc}
\end{equation}
The equation (\ref{eq:mu-tau_osc}) means that 
${
  s_3\simeq 0,
}$
which is consistent with the CHOOS results\cite{choos98}.

Once the the matter enhanced neutrino oscillations was proposed
\cite{mikheyev85}, 
it has soon been applied to the atmospheric neutrinos crossing 
the earth\cite{carlson86}. Recently much attention\cite{pantaleone98} 
has been devoted to this subject. However, it seems that no systmatic 
study has not been made yet, especially for the three flavor model of
neutrinos.    
 
The purpose of this paper is to reveal the conditions that amplify 
the effect of ${\theta_3}$ through matter oscillations so that we can 
determine this small mixing angle.  

\section{Matter Enhanced Atmospheric Neutrinos}

I assume that the masses of neutrinos are hierarchical, namely,
${m_1 <}$  ${m_2 < m_3}$, Furthermore I assume that 
the solar neutrino oscillations are attributed to 
${\delta m_{21}^2 \equiv m_2^2 -m_1^2}$. The atmospheric neutrino 
oscillations are derived by  ${\delta m_{32}^2 \equiv}$ ${ m_3^2 -m_2^2}$, 
the value of which is ${{\cal O}(10^{-3}) eV^2}$.
The solar neutrino problem may be solved by the matter enhanced, or just so 
oscillations. In either case, ${\delta m_{21}^2}$ is much smaller than 
${\delta m_{32}^2}$, and irrelevant to our concerns. In the following, we 
assign  ${\delta m_{21}^2 = 0}$, and 
${\Delta m^2 \equiv \delta m_{31}^2 = \delta m_{32}^2}$. 
The parameters contained in the  model are ${\theta_2}$, ${\theta_3}$,
and ${\Delta m^2}$.      

In calculating the matter effect of the earth, 
I use the density profile given by the ``Preliminary reference 
Earth model''(PREM)\cite{dziewonski81}. 

The prominent feature of our model is that the survival probability 
of electron neutrinos ${P(\nu_e \rightarrow \nu_e)}$ is completely 
determined by the three parameters, {\it i.e.}, ${\theta_3}$, 
${\Delta m^2/E}$ where $E$ is the neutrino energy, and the zenith angle $z$.
It is natural to define the resonance condition as
\begin{equation}
P(\nu_e \rightarrow \nu_e) =0
\label{eq:resonance}
\end{equation}
The resonance states form curves in the the three parameter space 
${(\theta_3,\ \Delta m^2/E,}$ ${\cos z)}$ corresponding to discretization of 
the resonance wavelength. 
I solve eq.(\ref{eq:resonance}) numerically, and
show in Fig.1 and Fig.2 the values of the three parameters at the resonance. 
Fig.1 and Fig.2 correspond to the longest and  the second longest wavelength, 
respectively.

The other probabilities ${P(\nu_e \rightarrow \nu_\mu)}$,
${P(\nu_e \rightarrow \nu_\tau)}$, ${P(\nu_\mu \rightarrow \nu_\mu)}$, and 
${P(\nu_\mu \rightarrow}$ ${ \nu_\tau)}$ also depend on the 
${\nu_\mu \leftrightarrow \nu_\tau}$ mixing angle ${\theta_2}$.

\section{Summary and Discussions}

I have investigated the resonance conditions of neutrinos passing through 
the earth in a wide range of parameters, although only a small mixing angle 
${\theta_3 < 13^{\circ}}$ has a practical meaning as has been suggested 
by the CHOOS experiments.

The results show that it is necessary to observe the high energy electron 
neutrinos crossing the depth of the earth to investgate the small 
mixing angle ${\theta_3}$.
For example, the resonance occurs at ${\theta_3 = 6.4^{\circ}}$, 
${\Delta m^2/E = 6.1\times 10^{-4} \ eV^2/GeV}$, and ${\cos z = -0.9}$.
The energy of electron neutrino is  ${5 \ GeV}$
if ${\Delta m^2 = 3 \times 10^{-3} eV^2}$.
This indicates the need to assemble the data of 
the zenith angle distribution of the electron neutrinos
whose energy is of order ${5 \ GeV}$ in order to determine the lepton 
mixing angle ${\theta_3}$.

This research was supported in part by the Grant-in-Aid for Scientific 
Research of the Ministry of Education, Scinece and Culture of Japan 
No. 10640281.

\vskip 5mm

\begin{flushleft}
{\Large \bf Figure Captions}
\end{flushleft}

\begin{description} 
\item Fig.1 The resonace conditions(I) as functions of ${\cos z}$. 
The solid line represents ${\theta_3}$, and the dot-dashed line represents 
${\Delta m^2/E}$.

\item  Fig.2 The resonace conditions(II) as functions of ${\cos z}$. 
The solid linerepresents ${\theta_3}$, and the dot-dashed line represents 
${\Delta m^2/E}$.
\end{description}

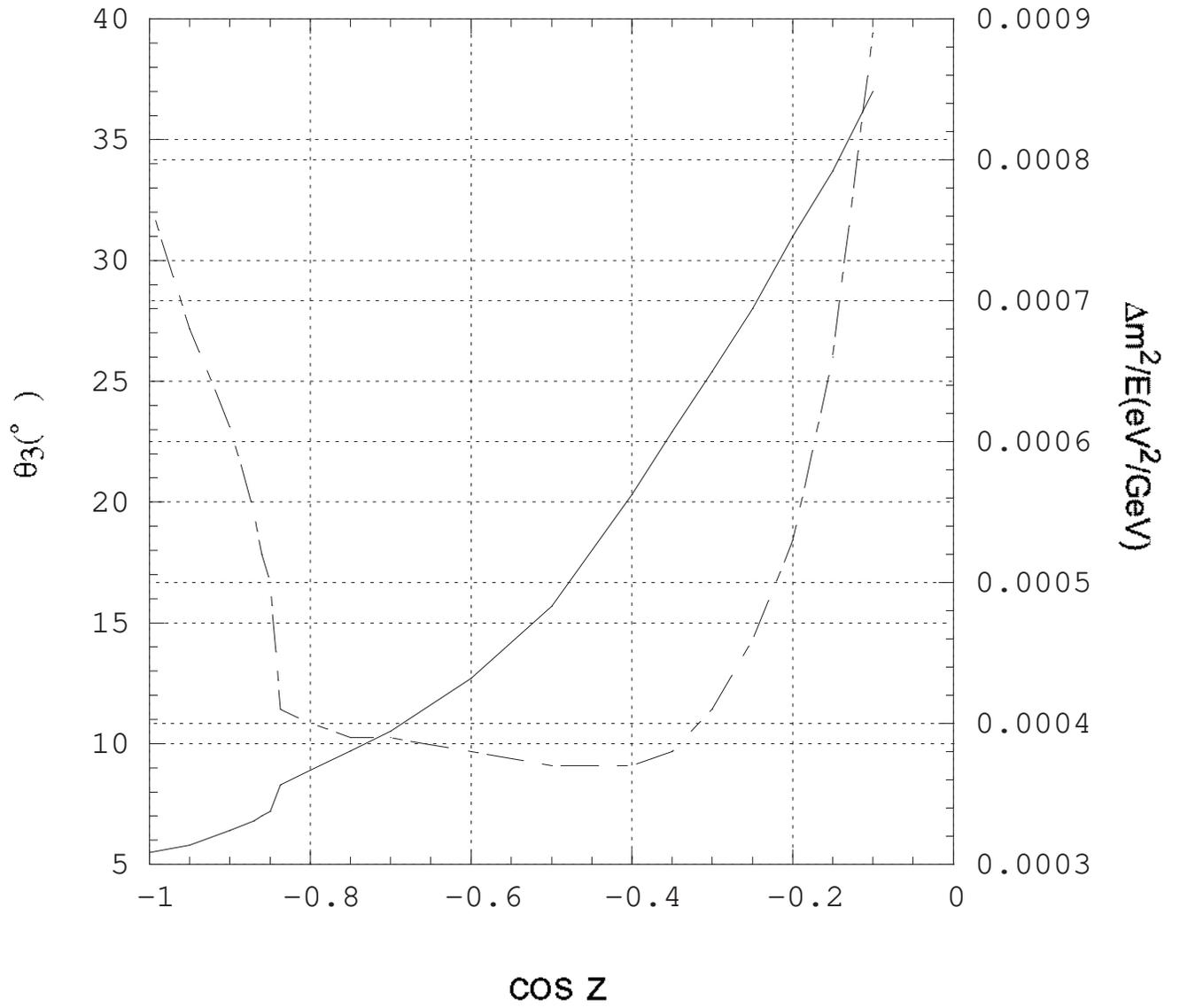
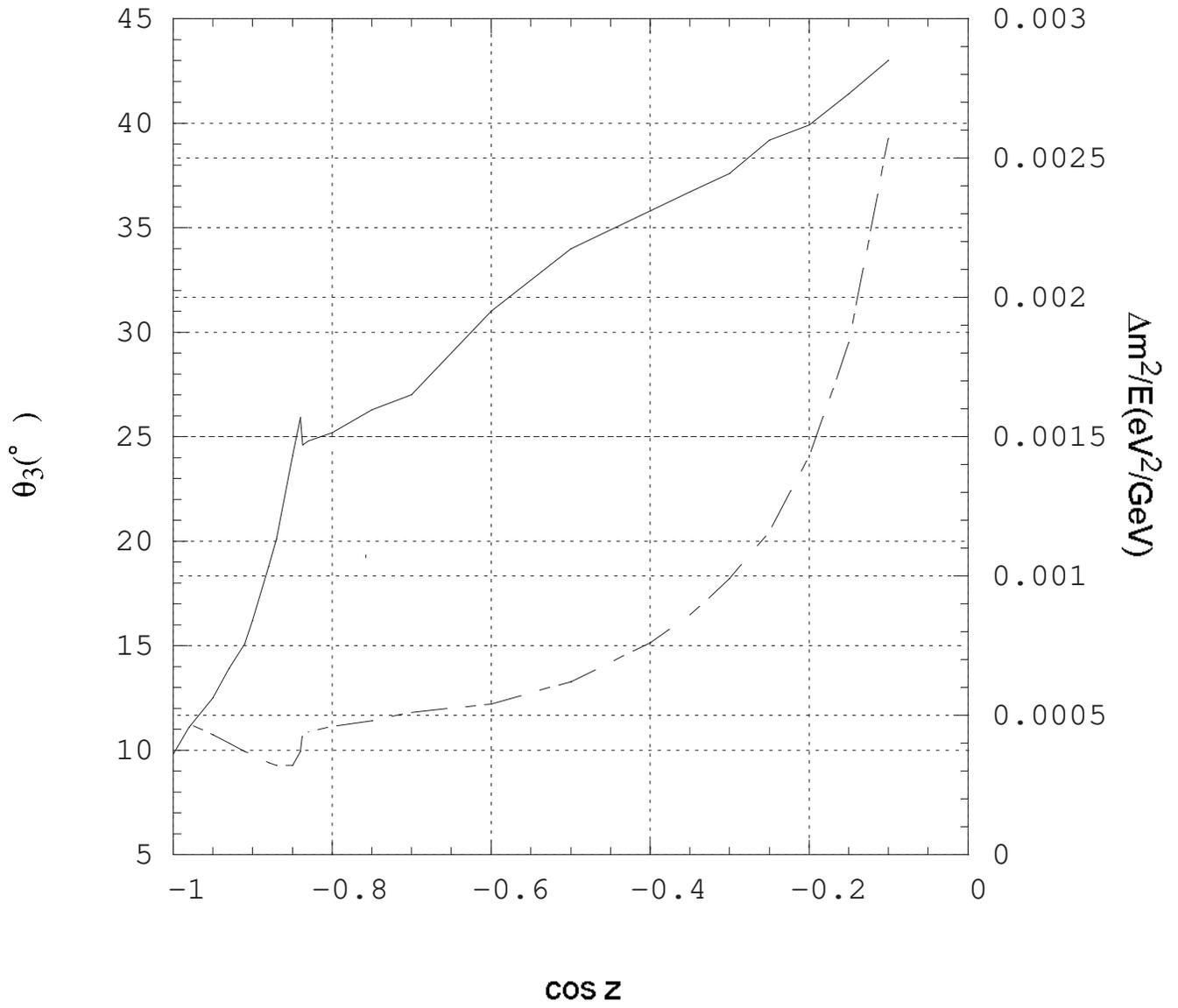
\end{document}